# Co$_2$MnZ (Z = Al, Si, Ga, Ge, Sn) Heusler alloys as candidate materials for spintronic and microelectronic applications: Electronic structure, transport, and magnetism

Vyacheslav V. Marchenkov[1,2*], Alena A. Semiannikova[1**], Evgenii D. Chernov[1],
Alexey V. Lukoyanov[1,2***], Valentin Yu. Irkhin[1], Yulia A. Perevozchikova[1],
Elena B. Marchenkova[1]

[1] M.N. Mikheev Institute of Metal Physics, UB RAS, 620108, S. Kovalevskoy 18, Yekaterinburg, Russia

[2] Ural Federal University, 620062, Mira 19, Yekaterinburg, Russia

*march@imp.uran.ru, ***lukoyanov@imp.uran.ru

**corresponding author: semiannikova@imp.uran.ru

**Abstract**

Magnetic and electronic transport properties of Co$_2$MnZ (Z = Al, Ga, Ge, Si, Sn) Heusler alloys were experimentally investigated. Electrical resistivity, in the temperature range from 4.2 to 300 K, as well as field dependences of the Hall effect and magnetization at $T$ = 4.2 K in magnetic fields of up to 100 kOe and 70 kOe, respectively, were measured. Experimental data are in good agreement with the results of the theoretical DFT calculations of the electronic structure and magnetic moments. In the band structure of Co$_2$MnSi, half-metallicity is formed with the full spin polarization and the half-metallic gap of about 0.6 eV. In Co$_2$MnZ (Z = Al, Ge, Sn), it is shifted from the Fermi energy by the hole pockets at point $\Gamma$, preventing thereby the formation of the half-metallic state. In a peculiar case of Co$_2$MnGa, the antisite defects are expected to determine structural and electronic properties. For the Co$_2$MnAl and Co$_2$MnGa topological semimetals, Weyl topological points are found at the Fermi energy; however, for Z = Si, Ge, Si, these features are located deeper within to the valence band. The results show that Co$_2$MnGe and Co$_2$MnSn are usual ferromagnets, Co$_2$MnAl and Co$_2$MnGa alloys are topological semimetals that can find application in microelectronics, while Co$_2$MnSi is a half-metallic ferromagnet that is in high demand in spintronics.

1. **Introduction**

The study of the physical properties of already known Heusler compounds are relevant although Heusler alloys were found more than a century ago [1]. Moreover, the search for and development of new alloys are still of great interest (see, e.g. [2–5]). The reason is unique functional properties and characteristics of Heusler alloys, including shape memory and magnetocaloric effects [6, 7], unusual states of half-metallic ferromagnets, spin gapless semiconductor [8–12], topological semimetal and insulator [13–17], etc.

Heusler alloys can be classified as follows. Full Heusler alloys are compounds with the $X_2YZ$ general formula; the half-Heusler alloys possess $XYZ$ formula; Heusler alloys with $XX'YZ$ formula are quaternary; $XX'YY'ZZ'$ Heusler alloys are called double-half, where $X$, $X'$, $Y$, and $Y'$ are usually transition metals; $Z$ and $Z'$ are $p$-elements of III–V groups in the periodic table [18, 19]. In addition, Heusler alloys consisting entirely of $d$-metals, which are called all-$d$-metal Heusler compounds [20, 21], and exotic compounds $Z_2XY$ [22, 23] have been found recently. Compounds with a high degree of current carriers spin polarization constitute a significant field in the series of Heusler alloys, since such materials are necessary to apply in spintronic devices [24]. Simultaneously, many of the full-Heusler alloys based on cobalt with $Co_2YZ$ formula exhibit the properties of a half-metallic ferromagnet [25, 26] with high spin polarization. A large amount of experimental, theoretical, and calculated data confirms the presence of features in the electronic structure, magnetic state, and electron transport of such materials [11, 12, 27–29]. Thus, the state of the half-metallic ferromagnetism is observed in $Co_2FeSi$ single crystals and $Co_2MnSi$ films, see the experimental works [30, 31]. The spin polarization coefficient $P = 93^{+7}_{-11}\%$ was measured in $Co_2MnSi$ films at room temperature [31]. These outstanding characteristics make $Co_2MnSi$ a vital component of numerous spintronic devices [24].

In addition, Co-based Heusler alloys can exhibit properties of topological semimetals. In [32], the electronic structure of $Co_2MnGa$ was studied using angle-resolved photoemission spectroscopy (ARPES), DFT calculations, and quantum transport. It was found that this compound is a topological semimetal, where Weyl fermion lines and topological surface states are observed. It was shown in [33] that the value of the anomalous Hall conductivity (AHC) can reach 1600 $\Omega$ cm at room temperature in $Co_2MnGa$ single crystal due to the topological features. In this case, the magnitude of the AHC in such topological materials can be manipulated by changing the Berry curvature. Since topological semimetals, especially Weyl semimetals, contain "massless" charge carriers (see, e.g., [11–13] and references therein), such materials are promising for ultrafast micro- and nanoelectronic devices.

Significant changes in the behavior of the magnetic and electronic properties of the $Co_2YZ$ alloys are observed with a change in the $Y$- and $Z$-components [25, 26]. In addition, certain correlations and patterns are observed between these properties; the reason is directly related to the electronic energy spectrum, namely, with the density of electronic states. Apparently, similar correlations could be observed in other electronic characteristics of $Co_2YZ$ compounds.

Although many Co$_2$YZ alloys — particularly the Co$_2$MnZ family — have been studied fairly extensively, both experimentally and theoretically (mostly treating different Z elements separately), a number of gaps remain. For example, Co$_2$MnSi is a half-metallic ferromagnet (HMF), whereas Co$_2$MnAl and Co$_2$MnGa exhibit properties of a topological semimetal (TSM). Therefore, it is important and interesting to trace the evolution of the electronic structure and, consequently, the changes in transport and magnetism across the transitions between HMF and TSM states, i.e., upon varying the Z component in Co$_2$YZ. This forms the basis of the present work.

This work is aimed at establishing the patterns of behavior and the relationship between the density of electronic states, magnetic, and electronic characteristics obtained from the theoretical calculations, electronic transport experiment, and magnetic measurements in the Co$_2$MnZ series of Heusler compounds, where Z = Al, Si, Ga, Ge, Sn.

## 2. Experimental and computational details

The Co$_2$MnZ (Z = Al, Si, Ga, Ge, Sn) Heusler alloys were synthesized by arc melting in a purified argon atmosphere, followed by annealing at 800°C for 48 h. Elemental and structural analysis of samples under study, magnetic properties and Hall effect measurements were performed at the Collective Access Center for Nanotechnologies and Advanced Materials at the Institute of Metal Physics, Ural Branch, Russian Academy of Sciences.

A local elemental analysis and mapping of large areas in characteristic radiation were performed using a TESCAN LMS scanning electron microscope in the secondary electrons (SE) and back scattered electrons (BSE) modes equipped with an energy dispersive spectroscopy (EDS) analyzer. X-ray phase analysis ($\theta/2\theta$) was carried out on a DRON-3 diffractometer in CuK$\alpha$ radiation in the angular range of 25°–120°. Electrical resistivity was measured by the standard four-probe technique on direct current with switching of electric current flowing through the sample. Magnetization measurements were performed at $T = 5$ K and in magnetic fields of up to 70 kOe using an PPMS-9 setup. An Oxford Instruments setup was used to measure the Hall effect at $T = 4.2$ K in magnetic fields of up to 100 kOe. Hall effect measurements were performed using the standard technique, describing in detail in [34]. The studied samples were in the shape of plates. The magnetic field vector was directed strictly perpendicular to the plane of the plates with an accuracy of ± 2 degrees (or ± 2.5%), and the electric current flowed along the largest face of the sample.

First-principles calculations were performed within framework of density functional theory (DFT). The exchange-correlation functional was chosen in generalized gradient approximation (GGA) form with PBE (Perdew-Burke-Enzerhof) pseudopotentials [35]. The electronic structure of the Co$_2$MnZ alloys was computed using the standard ultrasoft potentials from the pseudopotential library of Quantum ESPRESSO [36]. For modeling of the antisite defects, a supercell with 8 atoms (2 f.u.) was used, which is doubled in comparison with the other calculations for 4 atoms (1 f.u.). In all calculations, the plane-wave energy cutoff is taken as 60 Ry for the wave functions and 600 Ry for the charge density. For the calculations, the k-grid of $8 \times 8 \times 8$ points was used which was found sufficient to provide self-consistency of the calculations and to maintain the accuracy.

## 3. Results and discussion
### 3.1. Structure

The diffraction patterns of the compounds studied are presented in Figure 1. It is evident that the $Co_2MnAl$, $Co_2MnSi$, $Co_2MnGe$, and $Co_2MnSn$ Heusler alloys have superstructure reflections indicating the presence of long-range order, while the presence of reflections (111) and (200) specifies $L2_1$ ordering. The $Co_2MnGa$ alloy has only the main reflections, indicating the formation of a disordered structure of the $A2$ type (bcc) in the alloy. It should be noted that the peak intensity varies greatly in the $Co_2MnAl$, $Co_2MnSi$, $Co_2MnGe$, and $Co_2MnSn$ compounds, which may indicate incomplete ordering and/or small local deviations from stoichiometry.

Elemental analysis showed that the deviation from the stoichiometric composition is insignificant in all alloys, less than 2.5% for each element. Data on the structure, lattice constants, and elemental composition are presented in Table 1.

Calculations of the electronic structure and magnetic moments of the $Co_2MnZ$ ($Z$ = Al, Si, Ga, Ge, Sn) Heusler alloys were made using the obtained structural data (Table 1).

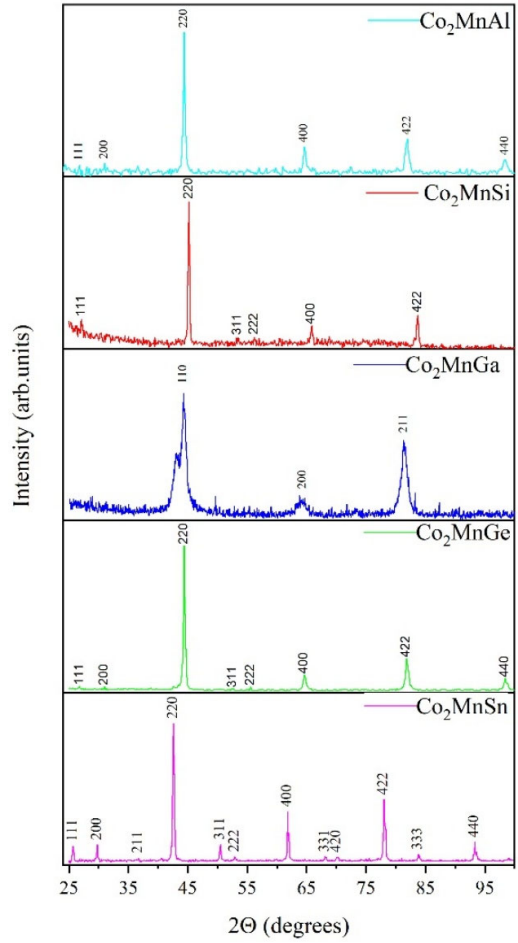

**Figure 1.** X-ray diffraction pattern of Co$_2$MnZ (Z = Al, Si, Ga, Ge, Sn) alloys at room temperature.

**Table 1.** Atomic number $z$, structure type, lattice constant $a$, and real composition of Co$_2$MnZ (Z = Al, Si, Ga, Ge, Sn) alloys

| Alloy | Atomic number $z$ | Structure | Lattice constant $a$, Å | Real composition |
|---|---|---|---|---|
| Co$_2$MnAl | 13 | $L2_1$ | 5.754 | Co$_2$Mn$_{1.1}$Al$_{0.9}$ |
| Co$_2$MnSi | 14 | $L2_1$ | 5.650 | Co$_{1.95}$MnSi$_{1.05}$ |
| Co$_2$MnGa | 31 | $A2$ | 2.880 | Co$_{1.9}$Mn$_{1.02}$Ga$_{1.08}$ |
| Co$_2$MnGe | 32 | $L2_1$ | 5.750 | Co$_{2.09}$Mn$_{1.01}$Ge$_{0.9}$ |
| Co$_2$MnSn | 50 | $L2_1$ | 5.987 | Co$_{1.96}$Mn$_{1.03}$Sn$_{1.01}$ |

## 3.2. DFT calculations: electronic structure and magnetic properties

Figure 2 shows the density of electronic states (DOS) of the $Co_2MnAl$. In the majority spin projection, $Co_2MnAl$ exhibits metallic properties with 1.56 st./eV/f.u. at the Fermi level. The Mn states are observed in the valence band with the peaks localized at –0.6 and –2.6 eV. The Co states are found in valence band as well, however, without pronounced peaks. In the minority spin projection, $Co_2MnAl$ is semi-metallic with the low DOS at the Fermi energy, i.e., 0.5 st./eV/f.u. at the Fermi level. The Mn states have a pronounced peak at 1.9 eV above zero. The Co states have two peaks in the valence and conductivity bands at –0.8 and 1.2 eV. The spin polarization is equal to 51%. The calculated $Co_2MnAl$ band structure is presented in Figure 2 with visible flat bands at the Fermi level. An energy gap in $Co_2MnAl$ is present above the hole pockets at point $\Gamma$ starting from 0.3 eV above the Fermi energy in the minority spin projection, and the shift of these bands prevents the occurrence of half-metallicity in $Co_2MnAl$. This implies that $Co_2MnAl$ exhibits topological semimetal properties according to [37] where the existence of Weyl points near the Fermi level was demonstrated. The $Co_2MnAl$ alloy has the total magnetic moment of 4.05 $\mu_B$ which is consistent with the experimental value of 4.07 $\mu_B$ [38]. The Co moment is 0.70 $\mu_B$, Mn has the higher magnetic moment 2.85 $\mu_B$, while the Al one is equal to –0.20 $\mu_B$.

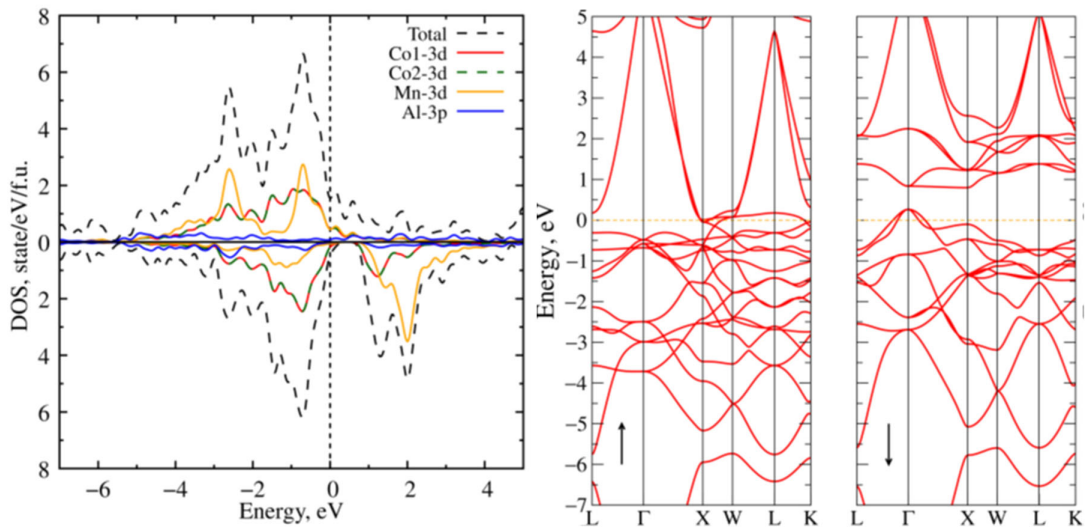

**Figure 2.** The density of electronic states (left) and band structure (right: majority and minority spin projections) of $Co_2MnAl$. The Fermi energy corresponds to zero energy indicated by a dotted line.

The DOS of $Co_2MnSi$ and the corresponding band structure are presented in Figure 3. A half-metallic gap with the width about 0.6 eV at the Fermi level for the minority spin projection can be seen in the right panel of Figure 3, so that $Co_2MnSi$ is a half-metallic ferromagnet (HMF). The Co states are strongly hybridized and form the narrow DOS peaks in the vicinity of the Fermi level for the minority spin projection at –1.2 and 0.8 eV. In addition, the Co and Mn states have the narrow peaks at –3.2 and –1.0 eV for the majority spin projection. The Mn 3d-states peaks are localized at the valence band for the majority spin projection and at 2 eV for the minority spin projection. Comparing the band structure of $Co_2MnAl$ (see Figure 3) with that of $Co_2MnSi$, it can be seen that when the *p*-element is replaced, the Weyl point and the hole pockets are shifted by –0.5 eV from the Fermi energy down to the valence band. $Co_2MnSi$ has the large total magnetic moment 5.00 $\mu_B$ nearly consistent with the experimental value 4.94 $\mu_B$. The Co magnetic moment is 0.95 $\mu_B$, 3.27 $\mu_B$ is for Mn, and the Si magnetic moment is as low as - 0.17 $\mu_B$.

The DOS and band structure of the next alloy $Co_2MnGa$ are presented in Figure 4. With changing of *p*-element to Ga, a shift of Co and Mn bands from valence band to the Fermi level is observed. It is worth paying attention to the intersection of three bands at high-symmetry point $X$ and the other intersection of two bands at high-symmetry point $W$, in the majority spin projection. Moreover, two flat bands between $X$ and $W$ and between $W$ and $K$ high-symmetry points are observed. All these facts point out to the presence of topology and qualify $Co_2MnGa$ as a Weyl topological semimetal (TSM) [32, 33, 39]. The number of electronic states at the Fermi level is equal to 1.8 st./eV/f.u. in the majority spin projection and 0.58 st./eV/f.u. in the minority spin projection, respectively. Thus, the spin polarization in $Co_2MnGa$ is calculated as 51%. Half-metallicity in $Co_2MnGa$ does not occur, mostly due to the bands in the minority spin projection with the hole pockets at the point $\Gamma$. The magnetic moments of ions are 0.73 $\mu_B$ for Co, 2.91 $\mu_B$ for Mn, and –0.15 $\mu_B$ for Ga. The total magnetic moment of $Co_2MnGa$ is obtained as 4.22 $\mu_B$ which is higher than experimental value 3.29 $\mu_B$. For some Heusler alloys, the reason for the discrepancy between theoretical and experimental values of magnetic moments was found to be atomic disorder [29].

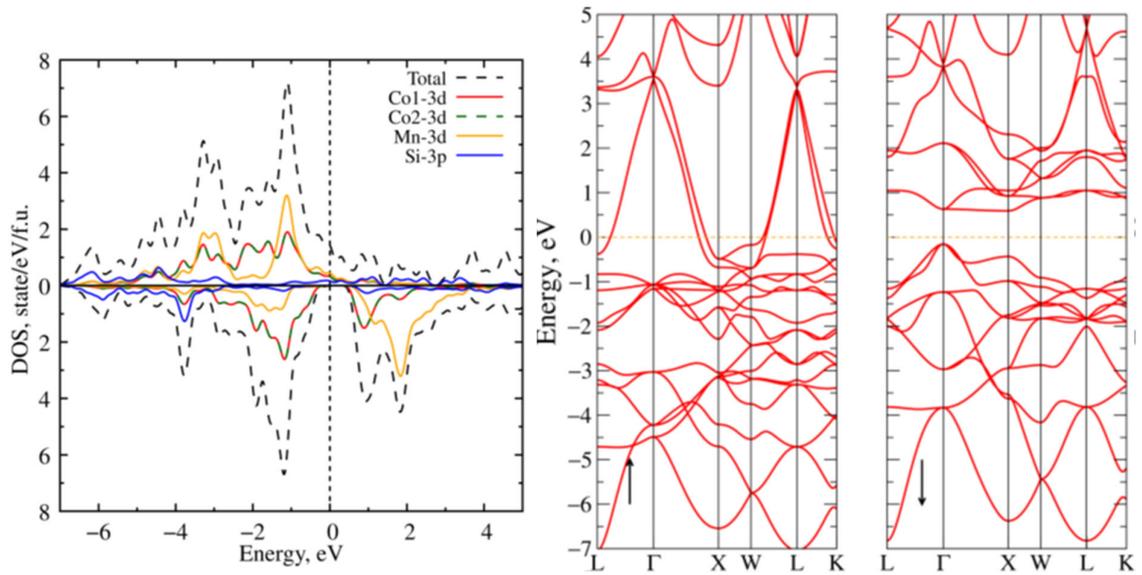

**Figure 3.** The density of electronic states (left) and band structure (right: majority and minority spin projections) of Co$_2$MnSi. The Fermi energy corresponds to zero energy indicated by a dotted line.

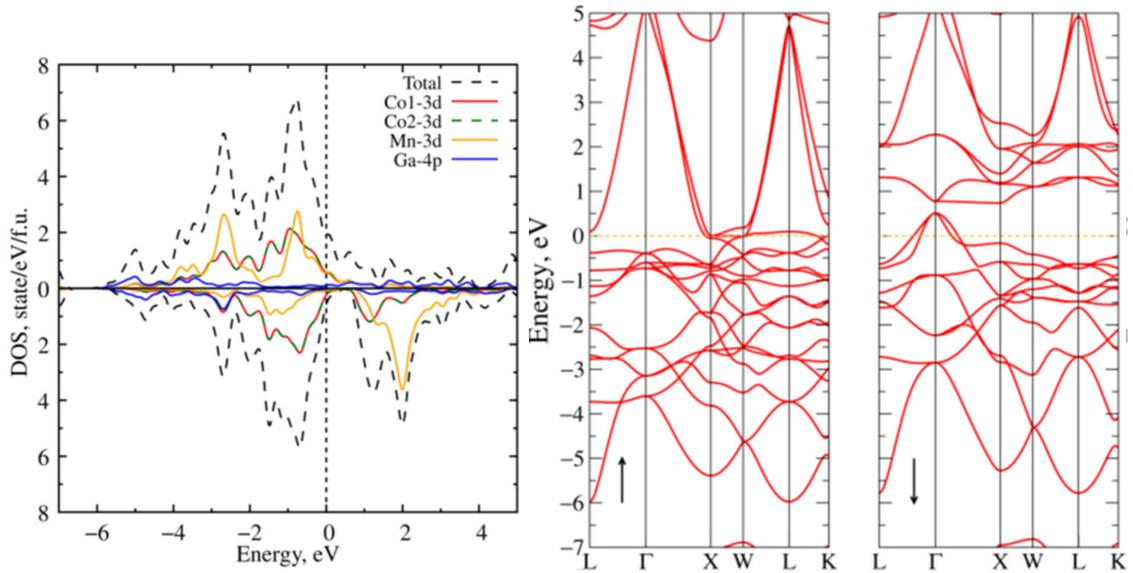

**Figure 4.** The density of electronic states (left) and band structure (right: majority and minority spin projections) of Co$_2$MnGa. The Fermi energy corresponds to zero energy indicated by a dotted line.

Figure 5 shows the calculation results for different types of the antisite defects in Co$_2$MnGa: (a) Co1-Mn; (b) Co2-Mn; (c) Co2-Ga: DOS changes dramatically. The band gap and strong hybridization of the Co states close. In addition, the intensity of individual state peaks increases greatly. As expected, when the Co ions are replaced by the Mn and Ga ions, the total magnetic moment changes. With the Co2-Ga replacement the total magnetic moment is 3.72 $\mu_B$. On the contrary, for the cases of the Co to Mn replacement (with the Co1-Mn and Co2-Mn cases resulting in the same magnetic moment), the value of the total magnetic moment increases, reaching 5.91 $\mu_B$.

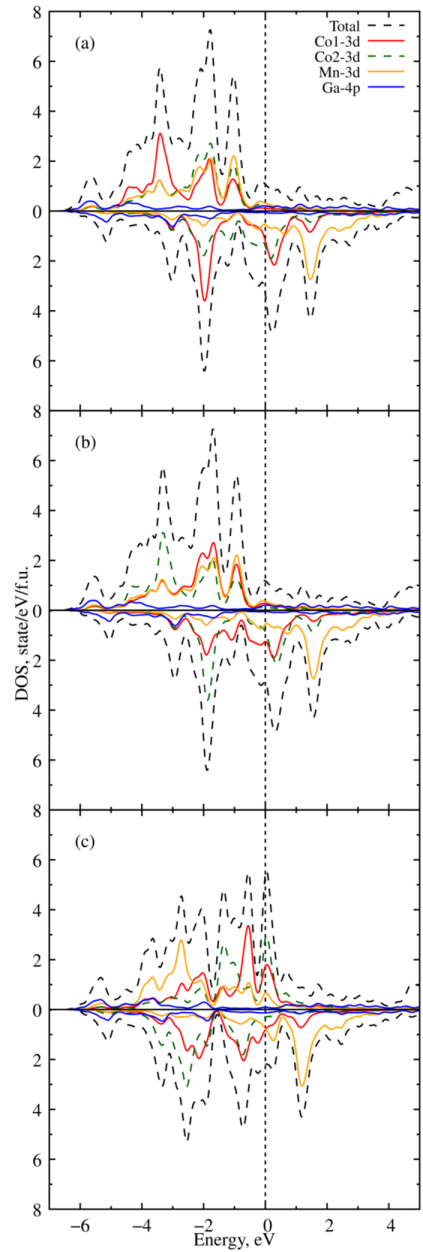

**Figure 5.** The density of electronic states of $Co_2MnGa$ with the antisite defects: (a) Co1-Mn; (b) Co2-Mn; (c) Co2-Ga. The Fermi energy corresponds to zero energy indicated by a vertical black dotted line.

The DOS and band structure presented in Figure 6 show that the Fermi level in the band structure of $Co_2MnGe$ lies near the gap edge in the minority spin projection. Compared to the previous alloy, in $Co_2MnGe$ the flat bands are shifted to valence band for the majority spin projection. The resulting value of spin polarization is equal to 92%. The $Co_2MnGe$ total magnetic moment is 5.01 $\mu_B$ which is consistent with the experimental value 5.00 $\mu_B$, the magnetic moments of individual ions are higher than in $Co_2MnGa$, namely, 3.13 $\mu_B$/Mn, 0.98 $\mu_B$/Co, and –0.08 $\mu_B$/Ge.

The DOS and band structure for $Co_2MnSn$ presented in Figure 7 are similar to those of $Co_2MnGe$. The number of electronic states at the Fermi level is equal to 1.49 st./eV/f.u. for the majority spin projection and 0.29 st./eV/f.u. for the minority spin projection while spin polarization is equal to 67% due to the hole pockets at the point $\Gamma$. The magnetic moments are 3.30 $\mu_B$ for Mn, 0.97 $\mu_B$ for Co, and –0.14 $\mu_B$ for Sn. These magnetic moments of individual ions are higher than in $Co_2MnGa$. The $Co_2MnSn$ total magnetic moment is 5.10 $\mu_B$ which is in agreement with the experimental value 4.90 $\mu_B$.

The values of the calculated magnetic moments $m_{cal}$, the density of electron states $N$ at the Fermi level $E_F$, and the degree of spin polarization $P$ obtained in our DFT calculations are presented in Table 2.

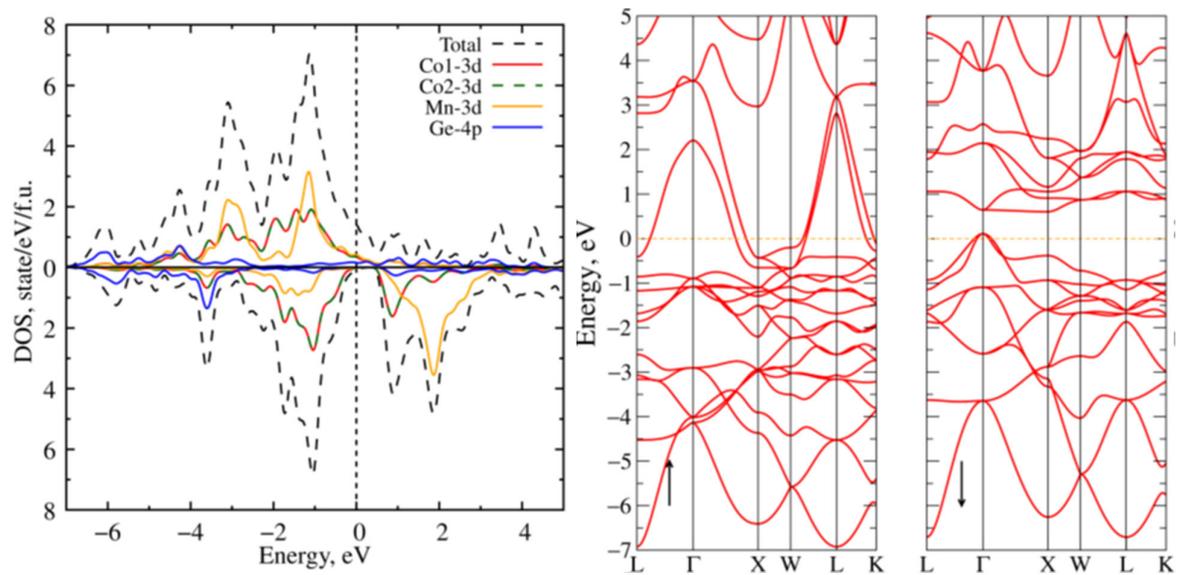

**Figure 6.** The density of electronic states (left) and band structure (right: majority and minority spin projections) of $Co_2MnGe$. The Fermi energy corresponds to zero energy indicated by a dotted line.

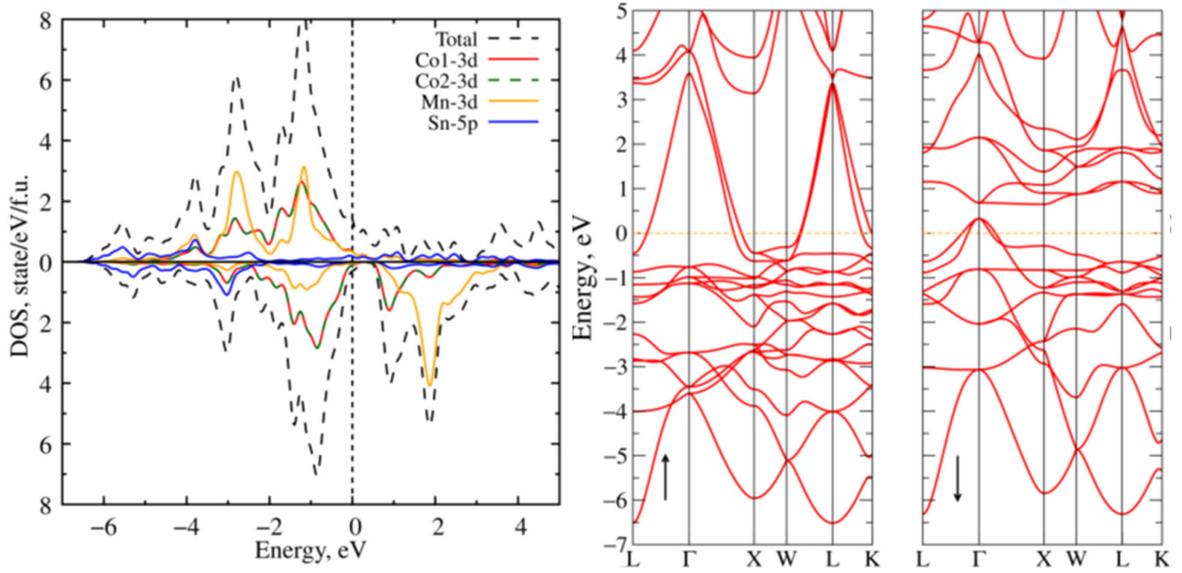

**Figure 7.** The density of electronic states (left) and band structure (right: majority and minority spin projections) of Co$_2$MnSn. The Fermi energy corresponds to zero energy indicated by a dotted line.

**Table 2.** Magnetic moments $m_{cal}$, the density of electron states $N$ at the Fermi level $E_F$, and the degree of spin polarization $P$ of Co$_2$MnZ ($Z$ = Al, Si, Ga, Ge, Sn). In addition, the values of magnetic moments determined experimentally $m_{s\ exp}$ and using the Slater-Pauling rule $m_{S-P}$ (see section 3.3) are presented.

| Compound | $m_{cal}$, $\mu_B$/f.u. | $m_{s\ exp}$, $\mu_B$/f.u. | $m_{S-P}$, $\mu_B$/f.u. | $N$, states / eV / f.u. | | | $P$ |
| --- | --- | --- | --- | --- | --- | --- | --- |
| | | | | up | down | total | |
| Co$_2$MnAl | 4.05 | 4.01 | 4 | 1.56 | 0.51 | 2.06 | 0.51 |
| Co$_2$MnSi | 5.00 | 4.94 | 5 | 1.45 | 1.00 | 1.45 | 1.00 |
| Co$_2$MnGa | 4.22 | 3.29 | 4 | 1.80 | 0.51 | 2.38 | 0.51 |
| Co$_2$MnGe | 5.01 | 5.00 | 5 | 1.41 | 0.92 | 1.47 | 0.92 |
| Co$_2$MnSn | 5.10 | 4.90 | 5 | 1.49 | 0.67 | 1.78 | 0.67 |

### 3.3. Magnetization and Hall effect

Figure 8a shows the field dependences of the magnetization $M(H)$ of Co$_2$MnZ ($Z$ = Al, Si, Ga, Ge, Sn) Heusler alloys at $T$ = 4.2 K. It is obvious that the magnetization of all the studied compounds reaches saturation in fields above 10 kOe (Fig. 8a). The experimental values of the magnetic moment $m_{s\ exp}$ in a field of 70 kOe at $T$ = 4.2 K are shown in Table 2. In Heusler alloys, the relationship between the magnetic moment $m_{S-P}$ and the number of valence electrons $N_{val.el.}$ is determined by the Slater-Pauling rule [40] and for Co$_2$-based alloys is written as

$$m_{S\text{-}P} = N_{val.el.} - 24 \qquad (1)$$

According to this rule, $m_{S\text{-}P} = 4$ for Co$_2$MnAl and Co$_2$MnGa alloys while $m_{S\text{-}P} = 5$ for Co$_2$MnSi, Co$_2$MnGe, and Co$_2$MnSn. The obtained experimental values $m_{s\ exp}$ correspond to the Slater-Pauling $m_{S\text{-}P}$ for all the compounds studied, except for the Co$_2$MnGa alloy with the magnetic moment $m_{s\ exp} = 3.29$ μ$_B$, which is significantly less than 4 μ$_B$ (see Table 2). Possible reasons for this discrepancy remain to be determined.

Figure 8b shows the field dependences of the Hall resistivity $\rho_{xy}(H)$ at $T = 4.2$ K. It is evident that the Hall resistivity $\rho_{xy}$ reaches saturation or a linear dependence in fields above 10–20 kOe as in the case of magnetization. Using the data on $M(H)$ and $\rho_{xy}(H)$, as well as the method described in [41], the normal $R_0$ and anomalous $R_s$ Hall coefficients were identified for all alloys. The corresponding data are presented in Table 3.

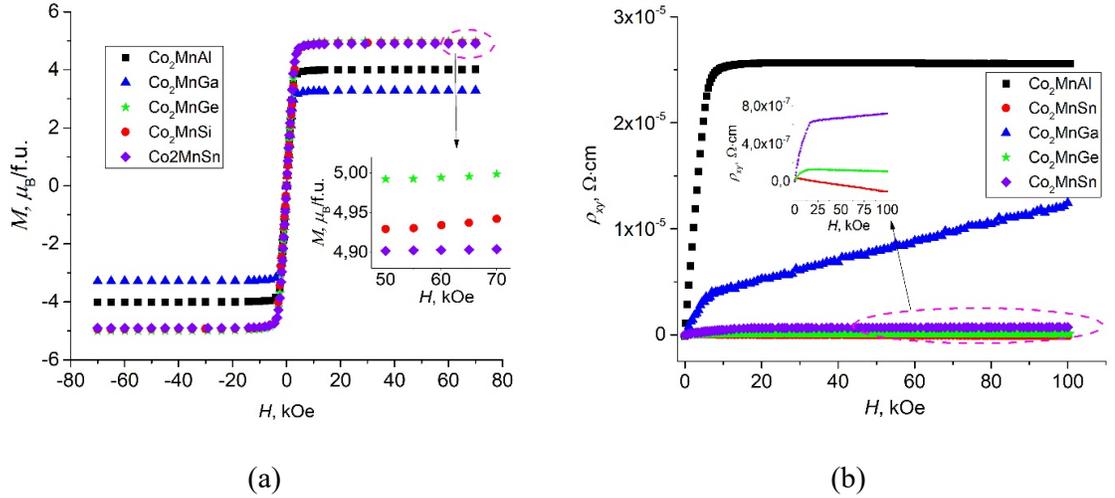

(a)          (b)

**Figure 8**. (a) Field dependences of the magnetization $M(H)$ at $T = 4.2$ K; (b) field dependences of the Hall resistivity $\rho_{xy}(H)$ at $T = 4.2$ K of Co$_2$Mn$Z$ ($Z$ = Al, Si, Ga, Ge, Sn) alloys.

It is evident (Table 3) that the value of the coefficient $R_s$ exceeds the value of $R_0$, as a rule, by two to three orders of magnitude, which is typical for ferromagnetic alloys. For all the alloys studied, regardless of the type of charge carriers, the coefficients of the anomalous Hall effect are positive. The compounds Co$_2$MnGa and Co$_2$MnSn have a positive coefficient of the normal Hall effect, i.e., the main charge carriers are holes. Among the alloys studied, the main charge carriers are electrons for Co$_2$MnAl, Co$_2$MnGe, and Co$_2$MnSi. The concentration and mobility of charge carriers was estimated using a single-band model [41], the results are given in Table 3.

**Table 3.** Normal $R_0$ and anomalous $R_s$ Hall coefficients, concentration $n$, and mobility $\mu$ of charge carriers of Co$_2$MnZ ($Z$ = Al, Si, Ga, Ge, Sn) alloys. The values of residual resistivity $\rho_0$ and coefficient $A$ at $T^2$ are also presented here (see Section 3.4).

| Alloy | Atomic number $z$ | $\rho_0$, $10^{-6}$ $\Omega$ cm | $A$, $10^{-10}$ $\Omega$ cm/K$^2$ | $R_0$, $10^{-4}$ cm$^3$/C | $R_s$, $10^{-2}$ cm$^3$/C | $n$, $10^{22}$ cm$^{-3}$ | $\mu$, cm$^2$/(V·s) |
|---|---|---|---|---|---|---|---|
| Co$_2$MnAl | 13 | 237.1 | - | -6.3 | 26.50 | 0.99 | 2.6 |
| Co$_2$MnSi | 14 | 15.5 | 2.43 | -1.4 | 0.04 | 4.50 | 9.1 |
| Co$_2$MnGa | 31 | 272.9 | - | 61.8 | 7.02 | 0.10 | 22.6 |
| Co$_2$MnGe | 32 | 23.4 | 5.22 | -0.7 | 0.15 | 8.90 | 3.0 |
| Co$_2$MnSn | 50 | 20.3 | 1.66 | 1.3 | 0.63 | 4.70 | 6.5 |

Table 3 shows that the current carrier concentration $n$ differs greatly for the two groups of compounds: for Co$_2$MnSi, Co$_2$MnGe, and Co$_2$MnSn alloys $n > 10^{22}$ cm$^{-3}$ while for Co$_2$MnAl and Co$_2$MnGa $n < 10^{22}$ cm$^{-3}$, i.e., significantly less.

### 3.4. Electrical resistivity

Figure 9a shows the temperature dependences of the specific electrical resistivity $\rho(T)$ of Co$_2$MnZ ($Z$ = Al, Ga, Ge, Si, Sn) alloys. Two types of alloys are observed that differ significantly both in the resistivity value and in the type of temperature dependence. Thus, for Co$_2$MnAl and Co$_2$MnGa alloys, the $\rho$ value is quite large, higher than 236 $\mu\Omega$ cm, and the $\rho(T)$ dependences have sections with a negative temperature coefficient of resistivity (TCR). Minima on $\rho(T)$ at $T = 48$ K for Co$_2$MnAl and $T = 28$ K for Co$_2$MnGa (Fig. 9a) and a dependence of $\sim T^{1/2}$ in the low-temperature region are observed (Figs. 9b). In the case of Co$_2$MnSi, Co$_2$MnGe, and Co$_2$MnSn alloys, the resistivity value is an order of magnitude smaller. The $\rho(T)$ dependence is "metallic", i.e., the resistivity increases monotonically with temperature. In addition, at temperatures up to 30 K, the resistivity changes according to a law close to a quadratic one, i.e., $\rho(T) = \rho_0 + A\,T^n$, $n \approx 2$, which is clearly seen in Fig. 9c. The values of the residual resistivity $\rho_0$ and the coefficient $A$, determined from the experiment, are given in Table 3.

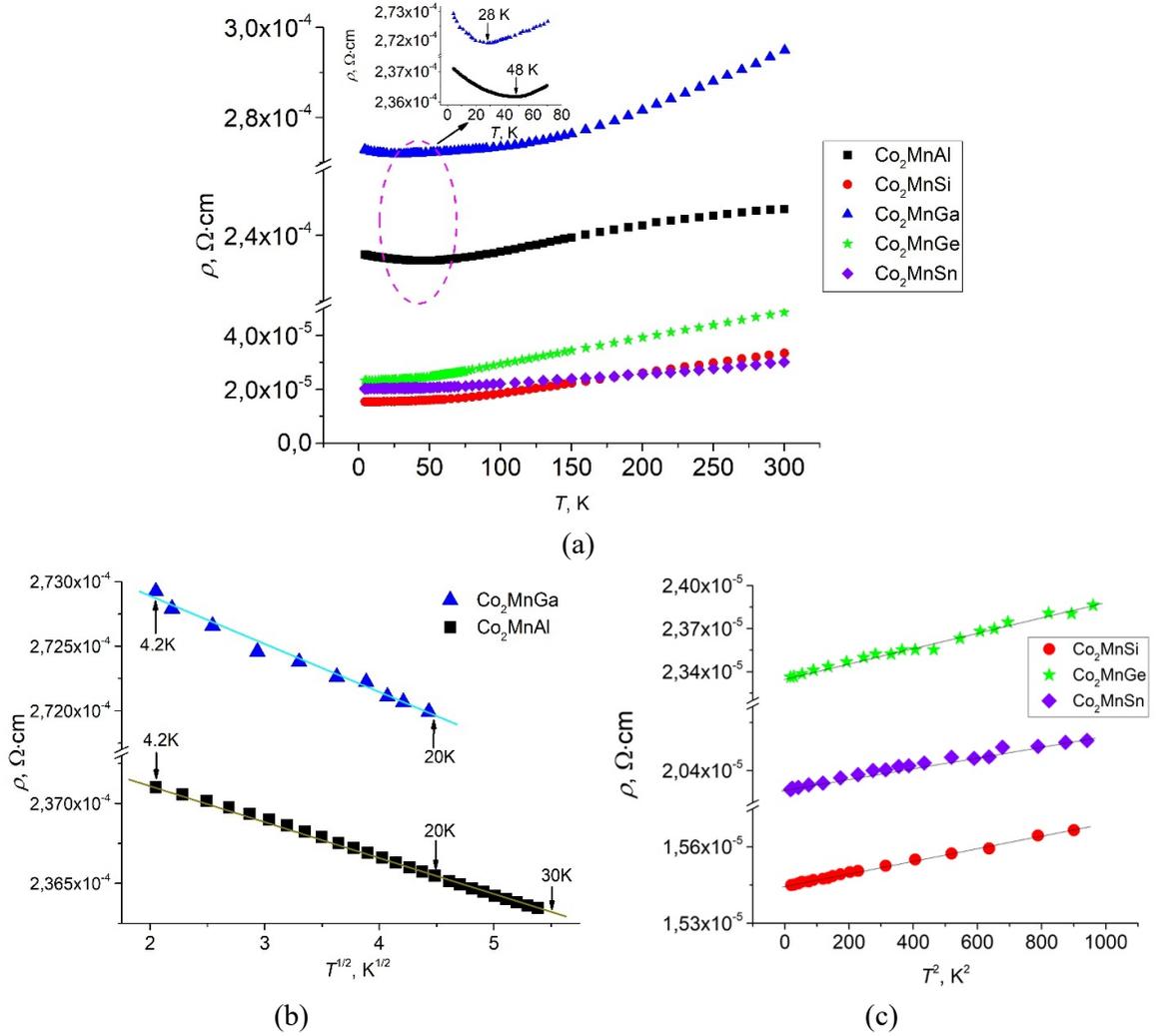

**Figure 9.** (a) Temperature dependences of the specific electrical resistivity $\rho(T)$; (b) dependences of $\sim T^{1/2}$ in the low-temperature region for Co$_2$MnAl and Co$_2$MnGa; (c) quadratic resistivity changes in the case of Co$_2$MnSi, Co$_2$MnGe, and Co$_2$MnSn alloys.

### 3.5. Electronic structure, electron transport, and magnetic characteristics interrelation

It is of interest to compare the electronic and magnetic characteristics obtained in the experiment with the data of DFT calculations of the electronic structure (section 3.2) for the Co$_2$Mn$Z$ ($Z$ = Al, Si, Ga, Ge, Sn) studied Heusler alloys.

Figure 10 shows the calculated values of the electron density of states $N$ at the Fermi level $E_F$ and the experimental values of the residual resistivity, the coefficients of the normal $R_0$ and anomalous Hall effect $R_s$ at 4.2 K depending on the atomic number $z$. A correlation between the calculated and experimental data is clearly observed, i.e., the maxima of the presented characteristics are found for the Co$_2$MnGa compound while the minima are evidenced for Co$_2$MnSi, Co$_2$MnSn, and "near" the Co$_2$MnGe (Fig. 10).

The dependences of the magnetic moment determined from the experiment $m_{s\,exp}$ and the calculated $m_{cal}$, as well as the calculated values of the spin polarization coefficient $P$ are illustrated in Figure 11. It is evident that the presented data $m_{s\,exp}$ and $m_{cal}$ are in good agreement with each other both in the magnitude of the magnetic moment and in the type of dependence on the atomic number $z$. The exception is the $Co_2MnGa$ alloy. In this case, the value of the magnetic moment is significantly less than the calculated one, as well as "determined" from the Slater-Pauling rule (1). In addition, a correlation is observed between the behavior of the magnetic moment and the calculated values of the spin polarization. Minima for $Co_2MnGa$ and maxima "near" $Co_2MnSi$ and $Co_2MnGe$ are observed. In this case, $P = 1$ for the $Co_2MnSi$ alloy.

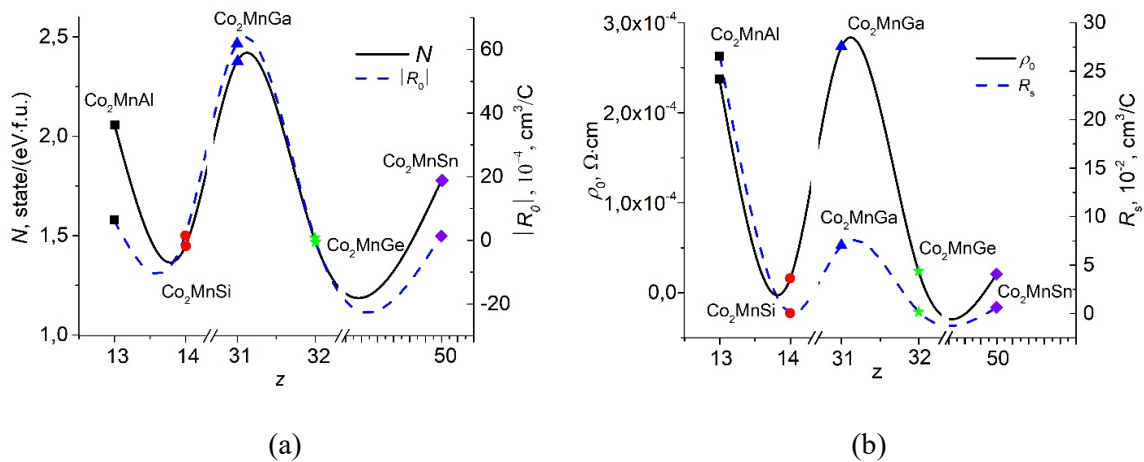

**Figure 10.** (a) Density of electron states $N$ at $E_F$ and coefficient (in absolute value) of the normal Hall effect $|R_0|$; (b) residual resistivity $\rho_0$ and anomalous Hall coefficient $R_s$ depending on the atomic number $z$.

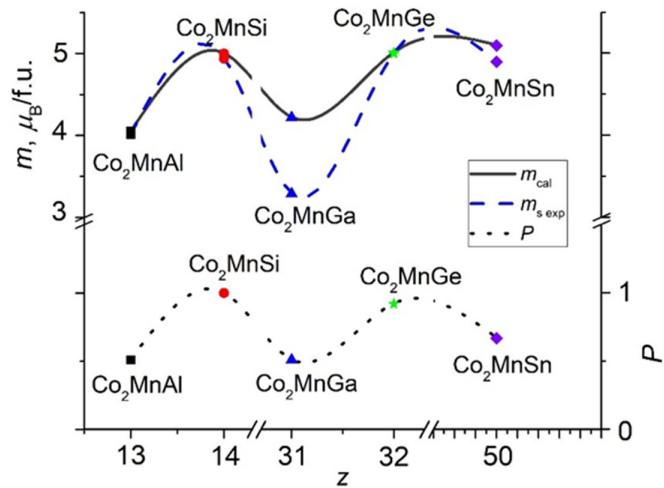

**Figure 11.** Experimental $m_{s\,exp}$ and calculated $m_{cal}$ magnetic moments and degree of spin polarization $P$ depending on the atomic number $z$.

Thus, the studied alloys can be divided into two types: 1) Co$_2$MnAl and Co$_2$MnGa alloys, exhibiting the properties of topological semimetals; 2) Co$_2$MnSi, Co$_2$MnGe, and Co$_2$MnSn ferromagnetic alloys.

### 3.5.1. Co$_2$MnAl and Co$_2$MnGa topological semimetals

The $\rho(T)$ dependences obtained (Fig. 9) are in many ways consistent with the works [42-43], where the Co$_2$MnAl and Co$_2$MnGa compounds were studied. Thus, in [42], $\rho(T)$ of epitaxial Co$_2$MnAl films was measured. It was found that the resistivity value is high exceeding 190 μΩ cm and the form of the $\rho(T)$ dependence is similar to our curve (Fig. 9a) with a resistivity minimum at $T \approx 50$ K [42]. In [43, 44], electron transport was studied, in particular, the temperature dependences of the resistivity of Co$_2$MnGa thin films of different thicknesses, and $\rho(T)$ dependences similar to ours with resistivity minima at low temperatures were observed. On the other hand, the data of the electron transport and magnetic properties of a Co$_2$MnGa single crystal measured in [45] are at variance with the presented results. In [45], the value of $\rho$ is smaller than in our case and varies from 60 to 130 μΩ cm, in the range from 50 K to room temperature. The smaller values of $\rho$ can be explained by the fact that a single-crystal sample was used in [45]. In addition, data on $\rho(T)$ at temperatures below 50 K omitted, where a low-temperature minimum may appear.

In our case, the high values of resistivity $\rho$ for Co$_2$MnAl and Co$_2$MnGa can be explained by the relatively low values of concentrations $n$ compared to Co$_2$MnSi, Co$_2$MnGe, and Co$_2$MnSn. Thus, in the case of Co$_2$MnAl $n \approx 9.9 \cdot 10^{21}$ cm$^{-3}$, and for Co$_2$MnGa it is even smaller, $n \approx 1 \cdot 10^{21}$ cm$^{-3}$, while for Co$_2$MnSi, Co$_2$MnGe, and Co$_2$MnSn alloys the current carrier concentrations are $4.5 \cdot 10^{22}$ cm$^{-3}$, $8.9 \cdot 10^{22}$ cm$^{-3}$, and $4.7 \cdot 10^{22}$ cm$^{-3}$, respectively.

Similar significant differences in the concentrations of charge carriers were observed in [46], where the electrical resistivity, magnetoresistivity, and Hall effect of amorphous and textured Co$_{1-x}$Si$_x$ ($0.4 < x < 0.6$) thin films were studied. It was shown that the type of temperature dependence $\rho(T)$ can change from metallic to semiconductor. In this case, the resistivity values $\rho$ are comparable or differ by no more than an order of magnitude. According to [46], a possible reason for such behavior of $\rho(T)$ is that Co$_{1-x}$Si$_x$ amorphous films have a lower mobility $\mu$, however, a higher concentration of charge carriers $n$, compared to textured Co$_{1-x}$Si$_x$ films. The difference in the concentration $n$ (one to two orders of magnitude) is apparently associated with a higher density of electron states in the HMF region at $x < 0.5$, and the strong difference in the values of carrier mobilities (one to two orders of magnitude) is due to structural disorder in amorphous films compared to structured ones [46]. Similar significant differences in the values of current carrier concentrations, as well as in their mobilities, are observed in our case.

Study of the electrical resistivity $\rho(T)$ of epitaxial $Co_2MnAl$ films with the $L2_1$ structure, in [47], found that at temperatures below 30 K, a contribution of $\rho \sim T^{1/2}$ is observed, caused by the interference of the electron-electron interaction and disorder, which leads to the emergence of a root singularity of the density of states at the Fermi level. This contribution is almost independent of the external magnetic field, however, closely related to the structural disorder. Similar dependences in the $Co_2MnAl$ and $Co_2MnGa$ alloys we observed at temperatures below 30 K and 20 K, respectively. This can indirectly confirm the presence of structural disorder, at least in $Co_2MnGa$.

It should be noted that in the case of $Co_2MnAl$ and $Co_2MnGa$, the value of the anomalous Hall coefficient $R_s$ is significantly higher than for the $Co_2MnSi$, $Co_2MnGe$, and $Co_2MnSn$ alloys and is equal to $R_s = 26.5 \cdot 10^{-2}$ cm$^3$/C for $Co_2MnAl$ and $R_s = 7.02 \cdot 10^{-2}$ cm$^3$/C for $Co_2MnGa$ (Table 3). A giant anomalous Hall effect, weakly dependent on temperature, was observed in $Co_2MnAl$ and $Co_2MnGa$ single crystals [15, 32, 33], where the authors associate the observed effects with the manifestation of the topological state in these materials. In our case, the large values of the anomalous Hall coefficient $R_s$ may indicate the manifestation of the topological semimetal state in the $Co_2MnAl$ and $Co_2MnGa$ alloys as well.

According to the Slater-Pauling rule (1), the magnetic moment for $Co_2MnAl$ and $Co_2MnGa$ should be $m_{S-P} = 4$ $\mu_B$. In the case of $Co_2MnAl$, this agrees well with our experimental and calculated data (Table 2, Fig. 11). For $Co_2MnGa$, a significant difference between rule (1) and calculation on the one hand, and experiment on the other was found, i.e., the experimental value $m_{s\;exp} = 3.29$ $\mu_B$ is significantly smaller than $m_{S-P} = 4$ $\mu_B$ and $m_{cal} = 4.22$ $\mu_B$. It was suggested above that a possible reason for such a discrepancy could be the presence of disorder, in particular, antisite defects. In this case (section 3.2), the total magnetic moment of the $Co_2MnGa$ alloy can vary from 3.72 $\mu_B$ to 5.91 $\mu_B$. As shown by X-ray structural analysis, disorder is indeed observed in the $Co_2MnGa$ alloy, one of the possible causes of which is the antisite defects of Co-Mn, Co-Ga.

### 3.5.2. $Co_2MnSi$, $Co_2MnGe$, and $Co_2MnSn$ ferromagnets

The values of residual resistivity of $Co_2MnSi$, $Co_2MnGe$, and $Co_2MnSn$ alloys are less than 24 $\mu\Omega$ cm, the temperature dependences $\rho(T)$ have a "metallic" appearance (Fig. 9a). The results we obtained on resistivity are in good agreement with many literature data (see, for example, [11, 25, 30] and references therein). The values of concentrations $n$ and mobilities $\mu$ of current carriers have values characteristic of polycrystalline metallic systems and lie within the range $n \approx (4.5 \div 8.9) \cdot 10^{22}$ cm$^{-3}$, $\mu \approx (3 \div 9.1)$ cm$^2$/V·s (Table 3).

The quadratic temperature dependence $\rho(T)$ observed in all Co$_2$MnSi, Co$_2$MnGe, and Co$_2$MnSn alloys at $T < 30$ K is apparently associated with electron-electron and electron-magnon scattering [48]. Then, in these alloys, the difference in the coefficients $A$ at $T^2$ should correlate with the concentrations of current carriers, as well as with the values of the coefficients of the anomalous Hall effect, if the skew scattering mechanism [49] is one of the main ones. In Table 3, the highest value of $A = 5.22 \cdot 10^{-4}$ µΩ cm·K$^{-2}$ among the studied alloys is observed in the Co$_2$MnGe alloy with the highest concentration value $n \approx 8.9 \cdot 10^{22}$ cm$^{-3}$, and a relatively large value of the anomalous Hall coefficient $R_s = 1.5 \cdot 10^{-3}$ cm$^3$/C.

The experimentally measured and calculated values of the magnetic moment of the Co$_2$MnSi, Co$_2$MnGe, and Co$_2$MnSn alloys are in good agreement with each other and with the value $m_{\text{S-P}} = 5$ µ$_B$ obtained using the Slater-Pauling rule (Table 3). The calculation data (Section 3.2) indicate a gap of about 0.6 eV in the density of electron states with spin down at the Fermi level $E_F$ for the Co$_2$MnSi alloy, which leads to 100% spin polarization $P$. In the case of the Co$_2$MnGe alloy, the DOS with spin down is slightly shifted toward the conduction band. As a result, the degree of polarization $P$ becomes less than 100%, however, it is still quite high and equal to 92%. Finally, in the Co$_2$MnSn alloy, the spin-down DOS is further shifted away from $E_F$, leading to a decrease in $P$ of up to 67%.

4. **Conclusions**

Thus, according to the experimental studies of electron transport and magnetic properties, as well as calculations of the electronic structure of Co$_2$Mn$Z$ ($Z$ = Al, Si, Ga, Ge, Sn) Heusler alloys, the following conclusions can be drawn.

The studied alloys can be divided into 2 groups: Co$_2$MnAl and Co$_2$MnGa compounds exhibit properties of topological semimetals, while Co$_2$MnSi, Co$_2$MnGe, and Co$_2$MnSn are ferromagnets. The Co$_2$MnAl and Co$_2$MnGa topological semimetals are characterized by high values of electrical resistivity > 236 µΩ cm, the presence of areas of negative TCR and minima on the temperature dependence of resistivity; low values of current carrier concentration ~ 10$^{21}$ cm$^{-3}$; relatively high value of the anomalous Hall coefficient ~ 10$^{-1}$ cm$^3$/C.

In the case of Co$_2$MnSi, Co$_2$MnGe, and Co$_2$MnSn ferromagnets, relatively small values of residual electrical resistivity ~ 15÷20 µΩ cm; "metallic" type of temperature dependence of resistivity and the value of current carrier concentration characteristic of metals ~ 10$^{22}$ cm$^{-3}$; comparatively small value of anomalous Hall coefficient ~ 10$^{-4}$÷10$^{-3}$ cm$^3$/C are observed.

The values of magnetic moments measured experimentally and obtained as a result of calculations are consistent with each other and obey the Slater-Pauling rule. The exception is the significantly smaller experimentally determined magnetic moment of the Co$_2$MnGa Heusler alloy. The observed difference can be explained by the presence of structural disorder in the form of antisite defects, which is confirmed by XRD and correlates with the calculated data.

Our theoretical calculations of the electronic structure allow us to reveal the differences in the nature of the ground state of the studied Heusler alloys in agreement with the experimental data, confirming the implementation of the topological semimetal state in the Co$_2$MnAl and Co$_2$MnGa alloys, conventional ferromagnetism in the Co$_2$MnSi, Co$_2$MnGe, and Co$_2$MnSn alloys, and half-metallic ferromagnetism in the Co$_2$MnSi alloy. As it clearly follows from the band structure analysis, half-metallicity in the other Co$_2$Mn$Z$ ($Z$ = Al, Ge, Sn) Heusler alloys vanishes due to the bands for the minority spin projection with the hole pockets at the point $\Gamma$; and by the expected antisite defects in Co$_2$MnGa as well. The Heusler alloys with the third group $p$-elements (Al an Ga) have topological Weyl points at the Fermi level, while for the fourth group $p$-elements (Si, Ge, Si), Weyl topological point is shifted to the valence band. The half- and semi-metallic properties of the Heusler alloys also depend on the choice of the $p$-element due to the presence of the hole pockets at the Fermi energy.

The manifestation of topological properties in the Co$_2$MnAl and Co$_2$MnGa alloys makes it possible to use them as materials for "fast" microelectronics, while the "half-metallicity" of the Co$_2$MnSi compound can be used in spintronics.

**Declaration of competing interest**

The authors declare that they have no known competing financial interests or personal relationships that could have appeared to influence the work reported in this paper.

**Acknowledgement**

The work was carried out within the framework of the state assignment of the Ministry of Science and Higher Education of the Russian Federation for the IMP UB RAS, and a part of research (study of Co$_2$MnAl and Section 3.5.1) was supported by Russian Science Foundation (project No. 24-72-00152, https://rscf.ru/project/24-72-00152/, M.N. Mikheev Institute of Metal Physics, Ural Branch of the Russian Academy of Sciences, Sverdlovsk region).

**Data availability**

Data will be made available on request.